\documentclass[english,letterpaper,pra,aps,showpacs,floatfix,twocolumn]{revtex4-1}
\usepackage[T1]{fontenc}
\usepackage[latin9]{inputenc}
\usepackage{verbatim}
\usepackage{color}
\usepackage{amssymb}
\usepackage{graphicx}

\makeatletter

\@ifundefined{textcolor}{}
{%
 \definecolor{BLACK}{gray}{0}
 \definecolor{WHITE}{gray}{1}
 \definecolor{RED}{rgb}{1,0,0}
 \definecolor{GREEN}{rgb}{0,1,0}
 \definecolor{BLUE}{rgb}{0,0,1}
 \definecolor{CYAN}{cmyk}{1,0,0,0}
 \definecolor{MAGENTA}{cmyk}{0,1,0,0}
 \definecolor{YELLOW}{cmyk}{0,0,1,0}
}

\begin{document}
\title{Quantum optomechanical straight-twin engine}


\author{Keye Zhang}
\email[Email address: ]{kyzhang@phy.ecnu.edu.cn}
\affiliation{Quantum Institute for Light and Atoms, School of Physics and Material Science, East China Normal University, Shanghai 200241, P.R. China}
\author{Weiping Zhang}
\affiliation{Department of Physics and Astronomy, Shanghai Jiao Tong University, Shanghai 200240, P.R. China}
\affiliation{Collaborative Innovation Center of Extreme Optics, Shanxi University, Taiyuan, Shanxi 030006, P. R. China}

\begin{abstract}
We propose a realization of a quantum heat engine in a hybrid microwave-optomechanical system that is the analog of a classical straight-twin engine. It exploits a pair of polariton modes that operate as out-of-phase quantum Otto cycles. A third polariton mode that is essential in the coupling of the optical and microwave fields is maintained in a quasi-dark mode to suppress disturbances from the mechanical noise. We also find that the fluctuations in the contributions to the total work from the two polariton modes are characterized by quantum correlations that generally lead to a reduction in the extractable work compared to its classical version.
\end{abstract}
\maketitle

\section{Introduction}
Research in quantum heat engines has progressed rapidly in recent years, due in part to its potential to investigate aspects of stochastic quantum thermodynamics at the microscopic and mesoscopic scales. This development has benefited in particular from emerging experimental technologies in AMO and NEMS systems, as evidenced in particular by the single ion heat engine theoretically proposed~\cite{Abah2012} and experimentally demonstrated~\cite{Rossnagel2016}. Additional approaches being considered  involve spin systems~\cite{Thomas2011, Altintas2015}, qubit systems~\cite{Brunner2012}, ultracold atoms~\cite{Fialko2012},  cavity-assisted atom-light  systems~\cite{Scully2003}, nanomechanical resonators \cite{Tercas2016}, and optomechanical systems~\cite{Zhang2014a, Zhang2014b, Mari2012, Mari2015, Elouard2015, Kurizki2015}. 

Quantum heat engines are typically based on implementations of quantum Carnot, Otto \cite{Quan2007}, Brayton \cite{Quan2009, Huang2013}, Diesel \cite{Quan2009, Dong2013} or Stirling cycles \cite{Wu1998, Huang2014}, with a succession of strokes that alternatively extract work and exchange heat with the hot and cold reservoirs. As a result the generation of work is an intermittent process. In classical thermodynamics, a transition to continuous work production can be achieved with multi-cylinder structures. In this paper we propose an implementation of a twin-cylinder optomechanical quantum heat engine. It is a hybrid microwave-opto-mechanical system with two working fluids -- or two ``cylinders''. The first one is a normal mode excitation (polariton) whose nature changes from microwave-like to optical-like, depending on the externally controlled position of the optomechanical oscillator (the analog of a piston) the engine being driven by the effective temperature difference between the microwave and the optical cavity modes. At the same time a second polariton mode that switches its nature in reversed sequence serves as a second working fluid that provides an additional work output channel operating out of phase with the first one. This is similar to the situation in classical straight-twin engines, but with the important difference that the work produced by the two polariton fluids can develop quantum correlations.

\begin{figure}[hbt]
  \includegraphics[width=8.5 cm]{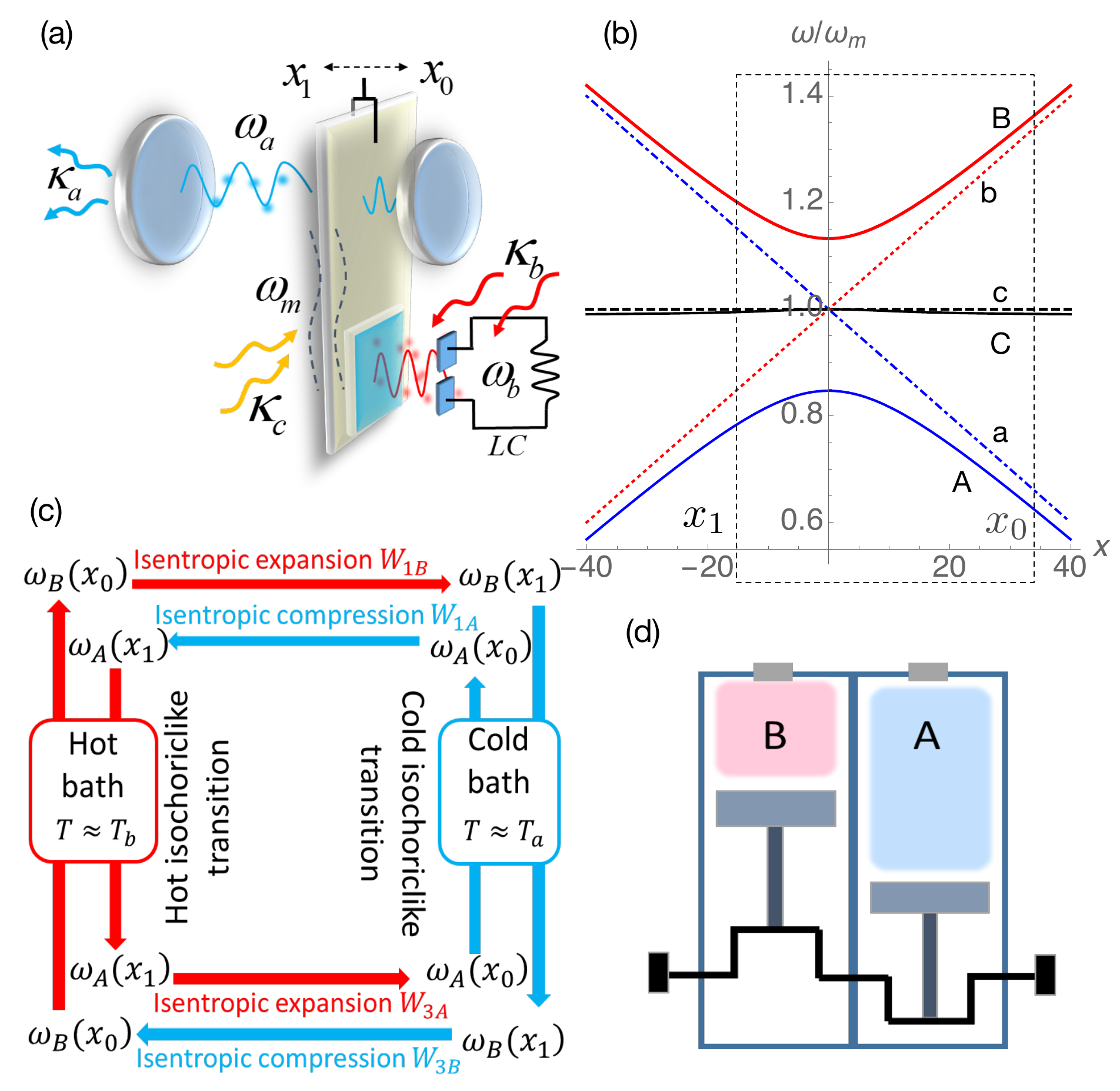}
  \caption{a) Quantum twin engine with an optomechanical mirror located between an optical cavity and a microwave resonator. (b) Polariton modes (solid) and bare modes (dashed) spectra as function of $x$ for $g_a=g_b=0.01\omega_m$, $G_b=-G_a=0.1\omega_m$, $\omega_{ap}-\omega_a=\omega_{bp}-\omega_b=-\omega_m$. The working range of the heat engine is indicated by the dotted rectangle. (c) Sketch of the Otto cycles associated with the polariton branches $B$ and $A$, respectively. (d) Sketch of an analog classical twin-cylinder heat engine.}
  \label{cycle}
\end{figure}

\section{Model}
As shown in Fig. \ref{cycle}(a) the engine consists of a pair of electromagnetic resonators that share a common optomechanical end mirror of oscillation frequency $\omega_m$ and damping rate $\kappa_c$. The first one is an optical cavity of frequency $\omega_a$ and damping rate $\kappa_a$ and the second one a microwave resonator of frequency $\omega_b$ and damping rate $\kappa_b$ \cite{Andrews2014}. They are driven by coherent fields at frequencies $\omega_{ap}$ and $\omega_{bp}$, respectively.  We assume that the intracavity fields are strong enough that they can be described as the sum of large mean classical amplitudes and small quantum fluctuations. In the frame rotating at the pump frequencies the system Hamiltonian can then be linearized as
\begin{equation}
H_1/\hbar=\omega_m \hat c^\dagger \hat c -\sum_{d=a,b}[\Delta_d \hat d^\dagger \hat d +G_d(\hat d+\hat d^\dagger)(\hat c+\hat c^\dagger)],
\end{equation}
where the bosonic annihilation operators $\hat a$, $\hat b$, and $\hat c$ account for the quantum fluctuations of the optical cavity field, microwave resonator field, and mechanical mode around their mean amplitudes $\langle \hat a\rangle$, $\langle \hat b \rangle$, and $\langle \hat c \rangle$, taken to be real without loss of generality. The effective optomechanical couplings, $G_a=-g_a \langle \hat a \rangle$, $G_b=g_b \langle \hat b \rangle$, and $g_a$ and $g_b$ are the single-photon optical and microwave couplings \cite{noteG}. In the following we assume that $G_a$ and $G_b$ are opposite in sign. Finally
\begin{equation}
\Delta_{a,b}(t) =\omega_{ap,bp}-\omega_{a,b}\pm g_{a,b}x(t),
\label{detuning}
\end{equation}
where we included the additional shift due to the classically controlled mean dimensionless displacement $x(t)=\langle \hat c+\hat c^\dagger\rangle$ of the compliant mirror, the ``$\pm$'' accounting for the implied opposite change in length of the two cavities.

The Hamiltonian $H_{1}$ can be diagonalized
in terms of three uncoupled bosonic normal modes, or polaritons, with
annihilation operators $\hat A$, $\hat B$ and $\hat C$ as
\begin{equation}
H_2=\hbar\omega_A \hat A^\dagger \hat A+\hbar\omega_B \hat B^\dagger \hat B+\hbar\omega_C\hat C^\dagger\hat C.
\end{equation}
The polariton modes are superpositions of the optical, microwave, and mechanical modes with relative amplitudes that depend on external parameters such as the detunings  $\Delta_{a,b}$. The fact that these constituents are coupled to thermal reservoirs at different temperatures allows for the realization of polaritonic heat engines as previously discussed in Refs.~\cite{Zhang2014a,Zhang2014b}. The present scheme is the simultaneously uses two polariton modes to realize a quantum analog of a classical two-cylinder straight-twin engine. 
  
The two ``cylinders'' of the engine are the polariton modes $A$ and $B$, whose nature and frequencies are controlled via $\Delta_{a,b}(t)$ by changing $x(t)$, the power source being the temperature difference between the microwave reservoir and the optical reservoir, which is effectively at zero temperature. Ideally the compliant mirror should merely allow for the conversion between microwave and optical photons so as to avoid unwanted energy losses through mechanical excitations. This suggests exploiting the ``polaritonic dark mode'' $\hat C=(-G_b \hat a+G_a \hat b)/\sqrt{G_a^2+G_b^2}$ and externally varying $G_{a,b}$ to achieve the adiabatic conversion between the microwave and optical photons free of the mechanical noise \cite{Wang2012a, Wang2012b, Tian2012, Dong2012}. However the realization of a perfect dark mode requires $-\Delta_a=-\Delta_b=\omega_m$, preventing the extraction of work in the adiabatic conversion.  But we find that if the detunings $\Delta_{a,b}(t) $  remain symmetric about the mechanical frequency $\omega_m$ throughout the thermodynamical cycle the polariton $C$ remains a quasi-dark  mode with nearly constant frequency $\omega_C \approx \omega_m$. This choice of detunings is optimal in keeping the mechanical excitations out of the heat-work conversion~\cite{note1}.

As an example we consider the case $\omega_{ap}-\omega_a=\omega_{bp}-\omega_b=-\omega_m$ and $g_a=g_b=g>0$, so that $\Delta_a=-\omega_m+gx$ and $\Delta_b=-\omega_m-gx$. Taking further $\omega_m\gg G_b=- G_a=G>0$ and dropping the anti-rotating terms in $H_1$ gives a three-mode Bogoliubov transformation 
\begin{eqnarray}
\hat A & = & \cos^2\frac{\theta}{2}\hat a-\sin^2\frac{\theta}{2}\hat b+\sqrt{2}\cos\frac{\theta}{2}\sin\frac{\theta}{2}\hat c,\nonumber\\
\hat B & = & -\sin^2\frac{\theta}{2}\hat a+\cos^2\frac{\theta}{2}\hat b+\sqrt{2}\cos\frac{\theta}{2}\sin\frac{\theta}{2}\hat c,\label{transform}\\
\hat C & = & \mp\frac{\sin\theta}{\sqrt{2}}(\hat a+\hat b)\pm\cos\theta\hat c, \nonumber
\end{eqnarray}
where $\theta=\arctan(\sqrt{2}G/gx)$ and the upper and lower signs in the third equation correspond to the positive and negative value of $x$, respectively. The polariton frequencies are $\omega_A=\omega_m-\sqrt{2G^2+g^2x^2}$, $\omega_B=\omega_m+\sqrt{2G^2+g^2x^2}$, and $\omega_C=\omega_m$. In the limit $gx \gg G$, Eqs. (\ref{transform}) give $\hat A\sim\hat a$ and $\hat B\sim \hat b$, while $\hat C\sim\hat c$ is approximately phonon-like. For $gx\ll-G$, $A$ and $B$ exchange their properties. 

Fig. \ref{cycle}(b) shows the polariton and bare modes frequencies as functions of the mean oscillator displacement $x$ for a symmetric arrangement of the optomechanical parameters of the optical and microwave fields. As $x$ is changed from large negative to large positive values the polaritons $A$ and $B$ switch their natures from microwave-like to optical-like and from optical-like to microwave-like, respectively, with an avoided crossing of frequency $2\sqrt{2}G$ at $x=0$. As already mentioned $C$ remains phonon-like. 

\section{Twin Otto cycle}
The heat engine is driven by the temperature difference between the microwave and optical reservoirs, with the polariton modes $A$ and $B$ undergoing asynchronous quantum Otto cycles. This is illustrated in Figs.~\ref{cycle}(c), which shows that when $A$ is in its isentropic compression stroke then $B$ is in the isentropic expansion stroke, and vice versa. This is analogous to the classical straight-twin-cylinder heat engine with a $\pi$ crankshaft angle of Fig.~\ref{cycle}(d), with one cylinder expanding with the other compressing.

We assume that initially $x=x_0$ and the system is in thermal equilibrium with the optical, microwave and phonon modes at the temperature of their respective reservoirs, with mean occupation numbers $\bar n_a$, $\bar n_b$ and $\bar n_c$.  For polariton $B$ the first stroke, an adiabatic change of $x$ from $x_0$ to $x_1$,  corresponds to a change from microwave-like with $\omega_{B}(x_0)\sim-\Delta_{b}(x_{0})$ and $\hat B\sim \hat b$ to optical-like. This step has to be fast enough that the interaction of the system with the thermal reservoirs can be largely neglected, yet slow enough that nonadiabatic transitions remain negligible. We then have $\bar n_B\sim \bar n_b$, so that if $x_0$ and $x_1$ are such that $\omega_{B}(x_{0})>\omega_{B}(x_1)$ that stroke is an isentropic expansion of polariton $B$, resulting in an energy loss and a work {\em output} $W_{1B}\approx\hbar[\omega_B(x_1)-\omega_B(x_0)]\bar n_b$. In contrast for polariton $A$ that stroke is an isentropic compression since $\omega_A(x_{0}) < \omega_A(x_1)$, resulting in an energy gain, hence a  work {\em input} $W_{1A}\approx\hbar[\omega_{A}(x_{1})-\omega_{A}(x_{0})]\bar n_a$. However its value is negligible since the temperature of the optical reservoir is effectively zero.

Following the second stroke, where the optical, microwave and phonon have reached thermal equilibrium again, the third stroke consists in adiabatically changing $x(t)$ back to $x_0$. Here the polariton $B$ suffers an isentropic compression, requiring a negligible work input, $W_{3B}\approx\hbar[\omega_{B}(x_{0})-\omega_{B}(x_{1})]\bar n_{a}\approx 0$, while $A$ undergoes an isentropic expansion with work output, $W_{3A}\approx\hbar[\omega_{A}(x_{0})-\omega_{A}(x_{1})]\bar n_b$. In the fourth and final stroke the system is left to reach thermal equilibrium again. Importantly during the full Otto cycle the polariton $C$ contributes negligible work since its frequency remains constant.  Ideally the total work output of the engine is therefore
\begin{eqnarray}
W_{\rm tot}&\approx& W_{1B} + W_{3A} \\
&\approx& \hbar[\omega_B(x_1)-\omega_A(x_1)+\omega_A(x_0)-\omega_B(x_0)]\bar n_b.	\nonumber \label{Wtot}
\end{eqnarray}
This simplified expression would indicate that maximizing the work requires an asymmetry and right-skewed working range with $x_0$ is large and positive and $x_1$ close to zero. However this choice results in an imperfect microwave-optical conversion of the polariton modes, the appearance of non-negligible contributions $W_{1A}$ and $W_{3B}$ to the total work, and the onset of quantum correlations between $W_{3A}$ and $W_{3B}$.  We return to this point later on. 

\section{Physical picture}
The operation of the engine can be understood in the bare mode picture by considering the combined effects of the radiation pressure forces from the quantum fluctuations of the microwave and optical fields and the classical control of the position $x(t)$ of the compliant mirror. Its dynamics can be described by a differential equation for the covariance matrix with Markovian-correlated quantum noise sources~\cite{Rogers2012}. 

\begin{figure}[hbt]
  \includegraphics[width=8.5 cm]{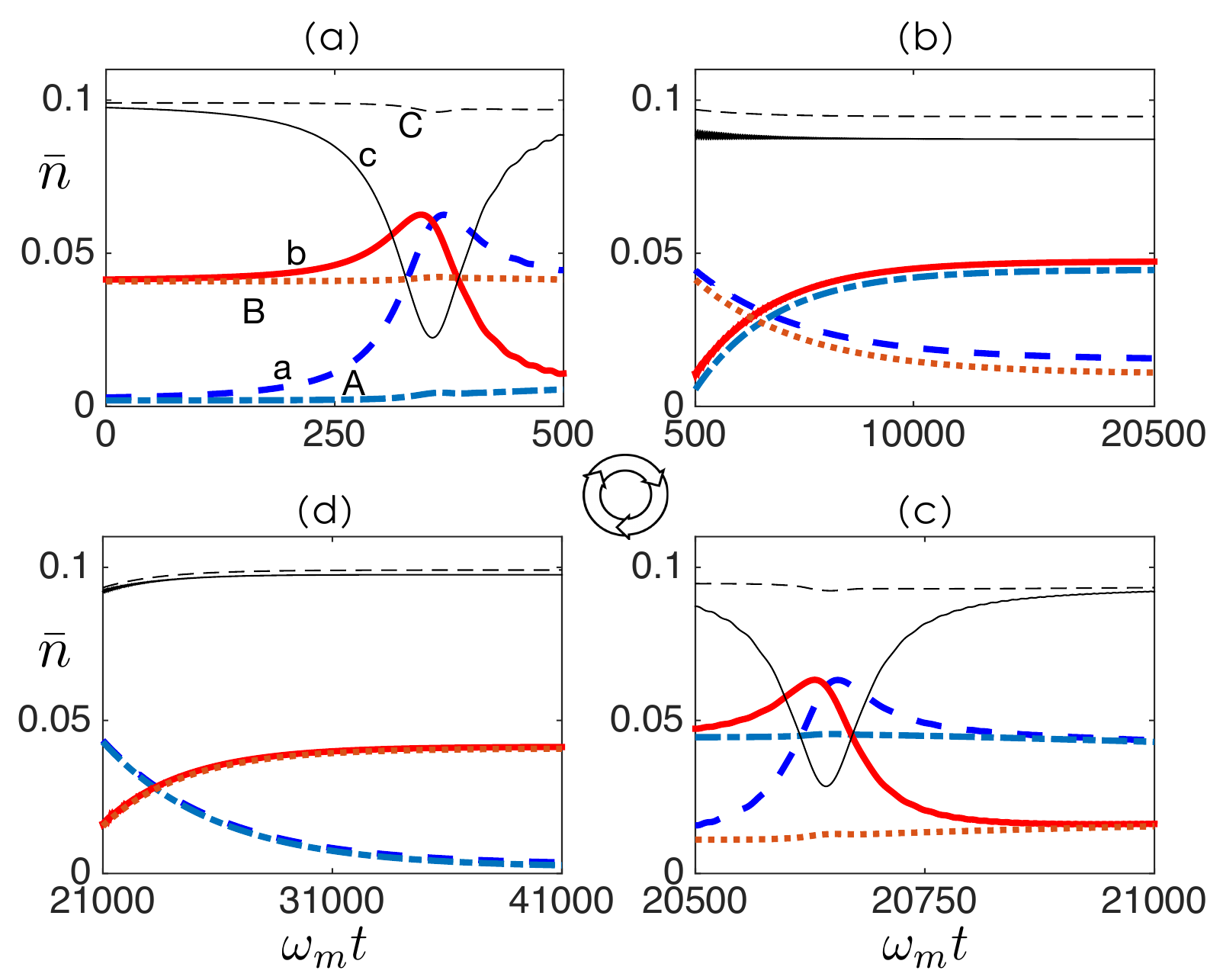}
  \caption{(color online) Evolution of the populations of the bare and polariton modes for one full cycle of the engine. The modes $a$ and $A$ are the blue dashed and dot-dashed lines, $b$ and $B$ the red solid and dotted lines, and $c$ and $C$ the thin black and dashed lines, respectively. The subplots (a)-(d) correspond to the strokes 1-4. Time in units of $\omega_M^{-1}$. The system parameters are given in the main text.}
  \label{phycycle}
\end{figure}

Fig.~\ref{phycycle} shows the evolution of the population of each mode for a full cycle of the engine obtained in this way. In this example the compliant mirror frequency is $\omega_m=2\pi\times 4$GHz and its temperature  $T_c=80$mK, resulting in $\bar n_c\approx 0.1$ \cite{Bochmann2013}.  For the optical cavity $\bar n_a\approx 0$ and for the microwave resonator $\bar n_b\approx0.04$. The linear optomechanical coupling is $G/\omega_m=0.1$ and the normalized optical, microwave and mechanical decay rates are $\kappa_{a,b,c}/\omega_m=0.0001$, $0.00014$, and $0.0002$, respectively.  In the first and third isentropic-like strokes the compliant mirror is moved linearly between $gx_0/\omega_m=1$ and $gx_1/\omega_m=-0.4$ and back. The duration of these strokes is  $\tau_{1,3}=500\omega_m^{-1}$ to satisfy the adiabatic requirement $G^{-1}\ll\tau_{1,3}\ll\kappa_{a,b,c}^{-1}$. The duration of the thermalizing second and fourth strokes is $\tau_{2,4}=2\times10^4\omega_m^{-1}>\kappa_{a,b}^{-1}$.

Initially the microwave and optical modes are thermally populated, but with the latter one is essentially empty due to the near-zero effective temperature of the optical reservoir. During the first stroke the compliant mirror is classically displaced from $x_0$ to $x_1$ and the radiation pressure force from the microwave photons produces work. (We choose the convention where work produced {\it by} the engine is negative.) For $x \approx 0$ the resonant condition $-\Delta_{a,b}=\omega_m$ is gradually approached and microwave photons convert into optical photons through the phonon-mediated interaction~\cite{Andrews2014, Bochmann2013} with a radiation pressure force acting against to the mirror motion, see Fig.~\ref{phycycle}(a). The asymmetry of $x_0$ and $x_1$ about $x=0$ results in that stroke ending before the positive work done by the microwave photons is offset by the negative work done by the optical photons. In the second stroke the optical field and microwave field rethermalize, again with $\bar n_a\approx 0$. In the third stroke the microwaves convert back to optical photons, see Fig.~\ref{phycycle}(c), and the optical radiation pressure force, which is along the displacement of the mirror in this stroke, becomes dominant and produces work. Finally the system is thermalized once more in stroke 4. 

Adiabatic conversion from the microwave to the optical mode occurs both in the first and the third stroke, as evidenced by the nearly constant population of the polariton modes $A$ and $B$. Importantly, as illustrated in Figs.~\ref{phycycle}(a) and \ref{phycycle}(c) the unequal distance of  $x_0$ and  $x_1$ from $x=0$ results in a larger initial detuning and a harder conversion and thus leaves a relatively long time for the microwave field to produce work. In contrast, during the third stroke the conversion occurs earlier so the population of the optical mode and the associated force are dominant most of the time. Note however that since $|x_1|$ is not quite large enough for the far-off-resonance condition to be fully  satisfied there is a non-negligible coupling between the three bare modes as they are thermalized in the second stroke. This results in a finite thermal equilibrium population of the optical mode and a small deviation between the evolution of the polaritons and the bare modes. This does not occur in the last stroke due to the far-off-resonant $x_0$. Except near resonance, where the phonons are directly involved in the optical-microwave conversion, their evolution coincides with the polariton mode $C$ whose population remains almost constant during the full engine cycle.

\section{Output work and correlation}
In the quantum adiabatic limit, the average work produced by polariton $A$ during stroke 3 is 
\begin{equation}
W_{3A} =  {\rm Tr}[\hat W_{3A}\hat \rho(x_1)]=\hbar[\omega_{A}(x_0)-\omega_{A}(x_1)]\bar n_B(x_1),
\label{W3A}
\end{equation}
where $\hat\rho(x_1)$ are the steady-state density matrix of the system and $\bar{n}_A(x_1)$ the population of polariton $A$ at $x=x_1$, and similarly for $W_{3B}$ with $A\rightarrow B$. The work produced during that stroke is $W_{3}=W_{3B}+W_{3A}$, and its variance ${\rm Var}[\hat W_{3}]={\rm Var}[\hat W_{3A}]+{\rm Var}[\hat W_{3B}]+2{\rm Cov}[\hat W_{3A},\hat W_{3B}]$. The non-classicality of the quantum twin engine is charactered by a nonzero correlation coefficient~\cite{Gerry2005}
\begin{equation}
J_{W_{3A},W_{3B}}=\frac{{\rm Cov}[\hat W_{3A},\hat W_{3B}]}{\sqrt{{\rm Var}[\hat W_{3A}]{\rm Var}[\hat W_{3B}]}}=-J_{n_A(x_1),n_B(x_1)}
\end{equation}
with $|J_{W_{3A},W_{3B}}| \le 1$. The expressions for stroke 1 are identical with $3\rightarrow 1$ and $x_1\rightarrow x_0$. 

To evaluate these quantities we introduce the quantum work operator $\hat W=\hat U^\dagger \hat H^\prime\hat U-\hat H$, where $\hat H$ and $\hat H^\prime$ are the Hamiltonians of the heat engine at the beginning and end of the isentropic strokes ~\cite{footnoteW}, and $\hat U$ is the time evolution operator
\begin{equation}
\hat U(t_f)=T_> \exp\left [ \frac{-i}{\hbar}\int_{t_i}^{t_f} dt \hat H(t) \right ]
\end{equation}
where $t_i$ and $t_f$ are the initial and final times of the stroke and $T_>$ the time-orderd product.

\begin{figure}[hbt]
\includegraphics[width=8.5 cm]{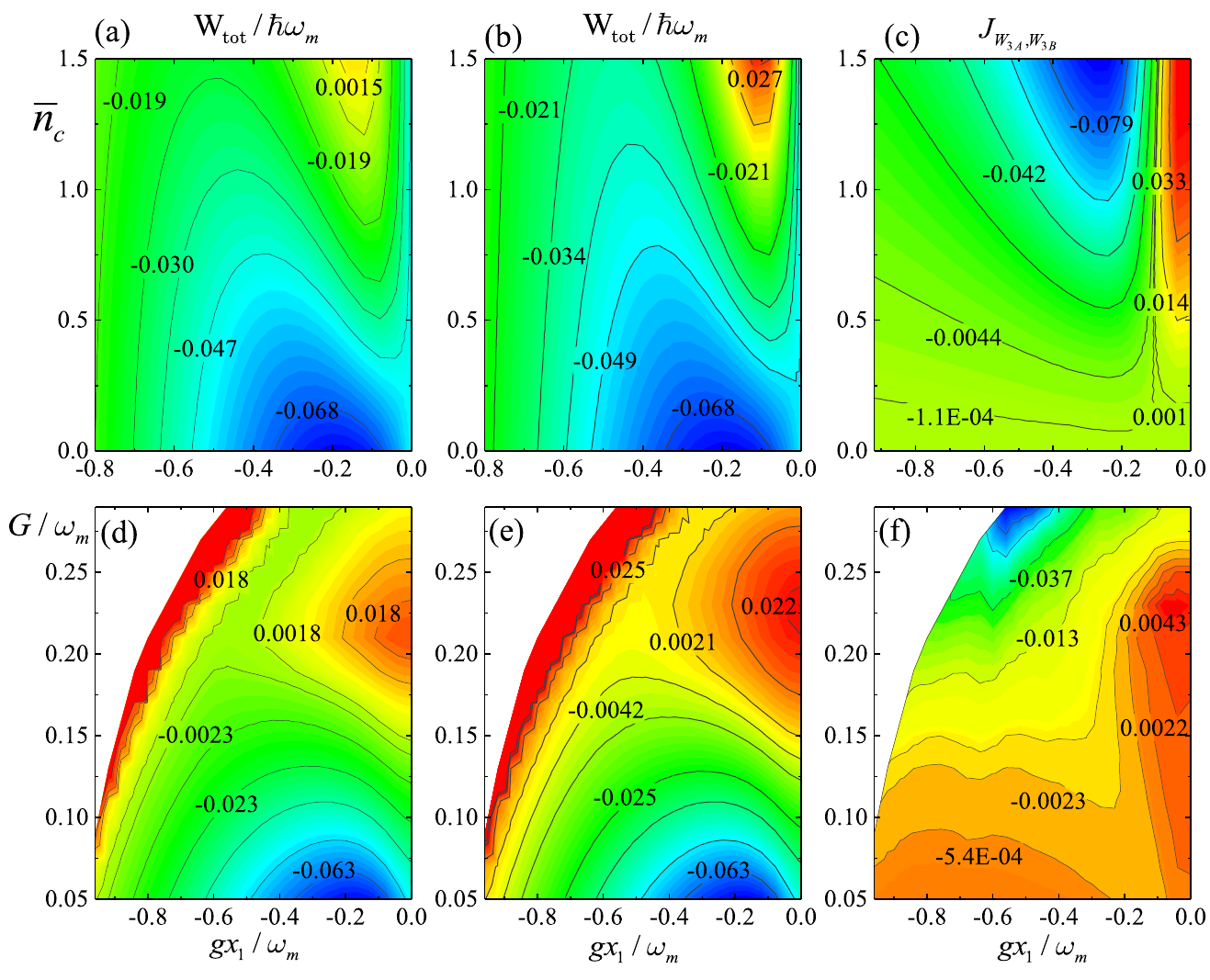}
\caption{Contour plots of the total work output (a) without and (b) with the classical approximation, and (c) the correlation coefficient $J_{W_{3A},W_{3B}}$ as a function of $x_1$ and the thermal mean phonon number $\bar n_c$ for $G/\omega_m$=0.1. Plots (d), (e) and (f): same, but as a function of the effective optomechanical coupling $G$ for $\bar n_c=0.2$. Other parameters as in Fig.~\ref{phycycle}(b). }
\label{worktot}
\end{figure}

For  $gx_0\gg G$ we have that $\bar n_B(x_0)\approx \bar n_b$, $\bar n_A(x_0)\approx \bar n_a \approx 0$, and $J_{W_{1A},W_{1B}}=-J_{n_A(x_0),n_B(x_0)} \approx 0$. With Eqs.~(\ref{transform}) and the associated polariton frequencies this gives
\begin{eqnarray}
W_{\rm tot} & \approx & \hbar(\sqrt{2G^2+g^2x_1^2}-\sqrt{2G^2+g^2x_0^2}) \nonumber\\
        &&\times[\bar n_A(x_1)-\bar n_B(x_1)+\bar n_b].
\end{eqnarray}  
The largest population difference, $[\bar n_A(x_1)-\bar n_B(x_1)+\bar n_b] = 2 \bar n_b$, occurs for $gx_1\sim -gx_0\ll-G$. However $(\sqrt{2G^2+g^2x_1^2}-\sqrt{2G^2+g^2x_0^2})=0$ in that limit, so that there is no net work. Extracting work from the heat engine requires therefore that neither $B$ nor $A$ be perfectly optical-like or microwave-like. Hence  the influence of the phonon mode and the resulting correlations between the $A$ and $B$ polaritons can not be ignored. 

The total work and quantum correlations can be estimated more accurately from the solution of the master equation in the polariton basis. In that basis the transformed Lindblad super-operators reveal the coupling of the polariton modes, implying effectively multi-mode correlated reservoirs (See the detailed derivation in the appendix). Fig.~\ref{worktot}(a) shows the dependence of the total work on $x_1$ on the mean thermal phonon number $\bar n_c$. The $\bar n_c$ dependence illustrates the detrimental effect of phonon reservoir temperature on $W_{\rm tot}$. This is because at higher temperature the optomechanical coupling increases the steady-state photon population, thereby decreasing the effective difference in temperatures of the polariton $B$ during the Otto cycle. For large $\bar n_c$ the total work can even change sign when $x_1$ is near $0$, as the polariton $B$ is then phonon-like rather than optical-like, and is warmer than the microwave-like side. Conversely, when $x_1$ is chosen large and negative the polariton $B$ becomes more and more optical-like, weakening the dependence of $W_{\rm tot}$ on $\bar n_c$. 

Figure~\ref{worktot}(b) shows for comparison a quasi-classical result obtained by neglecting the correlations. As in the quantum case the maximum work is reached near $gx_1=-0.2\omega_m$. The difference between the quantum and quasi-classical cases is most significant in the region of large $\bar n_c$ and small $x_1$, where the correlations between $A$ and $B$ become larger, resulting in decreased work. As shown in Fig.~\ref{worktot}(c), when $\bar n_c\gg\bar n_{a,b}$ the populations of $A$ and $B$ are both dominated by the thermalization of the phonon mode, resulting in negative work correlation. Note however that $J_{W_{3A},W_{3B}}$ becomes positive for $x_1\sim 0$ due to resonantly-enhanced conversion (the work correlation is opposite to polariton correlation).

A similar situation also occurs in the dependence of the work on the optomechanical coupling $G$, as shown in Figs.~\ref{worktot}(d)-(f), which illustrate the increase in work and decrease in quantum correlations as $G$ is decreased. Note finally that the validity of the linearized model imposes that $|gx|\ll\sqrt{\omega_m^2-8G^2}$. The positive $W_{\rm tot}$ and negative work correlation near this boundary are caused by the anti-rotating terms in the optomechanical coupling \cite{noteRWA}.

\section{Conclusion}
To summarize, we have proposed and analyzed a quantum optomechanical heat engine driven by the effective temperature difference between microwave and the optical fields, and with a classical-like twin-cylinder four-stroke structure that dramatically decreases the intervals between work extractions in the quantum Otto cycle and brings quantum correlation between work from each cylinder. Future work will extend this idea to structures with more cylinders, include an autonomous engine design~\cite{Garcia2016, Roulet2016}, and explore the applications of work correlation.

\acknowledgements
We acknowledge enlightening discussions with  P. Meystre, S. Singh, H. Pu and L. Zhou. KZ is supported by NSFC Grants No. 11574086 and No. 11654005, and the Shanghai Rising-Star Program 16QA1401600. WZ is supported by the National Key Research Program of China under Grant No. 2016YFA0302001 and NSFC Grants No.~11234003 and No.~11129402. 

\appendix*
\section{The master equation in the polariton basis}
The master equation in the bare mode basis writes
\begin{equation}
\frac{d\rho}{dt}=-\frac{i}{\hbar}[H_1,\rho]+\sum_{o=a,b,c}[\kappa_o(\bar n_o+1)\mathcal{L}_{\hat o}+\kappa_o\bar n_o\mathcal{L}_{\hat o^\dagger}]\rho,
\end{equation}
where $\kappa_o$ and $\bar n_o$ are the decay rate and the mean thermal occupation number of bare mode $o$, $H_1$ is the linearized optomechanical Hamiltonian
\begin{equation}
H_1/\hbar=\omega_m \hat c^\dagger \hat c -\sum_{o=a,b}[\Delta_o \hat o^\dagger \hat o +G(\hat o+\hat o^\dagger)(\hat c+\hat c^\dagger)],
\end{equation}
and the Lindblad superoperators are
\begin{equation}
\mathcal{L}_{\hat o}\rho=\hat o\rho\hat o^\dagger-\frac{1}{2}\hat o^\dagger\hat o\rho-\frac{1}{2}\rho\hat o^\dagger\hat o.
\end{equation}
When $G\ll\omega_m$, the analytical form of the inverse Bogoliubov transformation can be derived under the rotating wave approximation (RWA), which is
\begin{eqnarray}
\hat a & = & \cos^2\frac{\theta}{2}\hat A-\sin^2\frac{\theta}{2}\hat B\mp\sqrt{2}\cos\frac{\theta}{2}\sin\frac{\theta}{2}\hat C,\nonumber\\
\hat b & = & -\sin^2\frac{\theta}{2}\hat A+\cos^2\frac{\theta}{2}\hat B\mp\sqrt{2}\cos\frac{\theta}{2}\sin\frac{\theta}{2}\hat C,\label{transform2}\nonumber\\
\hat c & = & \frac{\sin\theta}{\sqrt{2}}(\hat A+\hat B)\pm\cos\theta\hat C. 
\end{eqnarray}
Substituting it into Eq. (A.1) we can rewrite the master equation in polariton basis,
\begin{eqnarray}
\frac{d\rho}{dt} & = & \sum_{j}-i\omega_j[\hat j^\dagger\hat j,\rho]+[\kappa_j(\bar n_j+1)\mathcal{L}_{\hat j}+\kappa_j\bar n_j\mathcal{L}_{\hat o^\dagger}]\rho \nonumber\\
& &+\sum_{jk}[(R_{jk}+S_{jk})\mathcal{J}_{\hat j\hat k^\dagger}\rho+S_{jk}\mathcal{J}_{\hat j^\dagger \hat k}\rho+h.c.] ,\nonumber\\
\label{masterABC}
\end{eqnarray}
where $j=A,B,C$ and $jk=AB, AC, BC$. In addition to the ordinary dissipation terms with effective polariton decay rates and mean thermal occupation numbers  
\begin{eqnarray}
\kappa_{A,B} & = & \kappa_{a,b}\cos^4\frac{\theta}{2}+\kappa_{b,a}\sin^4\frac{\theta}{2}+\kappa_c\frac{\sin^2\theta}{2},\nonumber\\
\bar n_{A,B} & = &(\kappa_{a,b}\bar n_{a,b}\cos^4\frac{\theta}{2}+\kappa_{b,a}\bar n_{b,a}\sin^4\frac{\theta}{2}\nonumber\\
&&+\kappa_c\bar n_c\frac{\sin^2\theta}{2})/\kappa_{A,B}, \\
\kappa_C &=& \frac{\kappa_a+\kappa_b}{2}\sin^2\theta+\kappa_c\cos^2\theta,\nonumber\\
\bar n_C &=& (\frac{\kappa_a\bar n_a+\kappa_b\bar n_b}{2}\sin^2\theta+\kappa_c\bar n_c\cos^2\theta)/\kappa_C,\nonumber
\end{eqnarray}
the transformed Lindblad superoperators of the bare modes bring out several coupling terms of the polariton modes in the form of the superoperators
\begin{equation}
\mathcal{J}_{\hat j\hat k^\dagger}\rho=\hat j\rho\hat k^\dagger-\frac{1}{2}\hat k^\dagger\hat j\rho-\frac{1}{2}\rho\hat k^\dagger\hat j,
\end{equation}
with the coupling coefficients
\begin{eqnarray}
S_{AB} & = & (2\kappa_c-\kappa_a-\kappa_b)\frac{\sin^2\theta}{4},\\
S_{AC,BC} & = &\frac{\sin\theta}{\sqrt{2}}(\kappa_{b,a}\sin^2\frac{\theta}{2}-\kappa_{a,b}\cos^2\frac{\theta}{2}\nonumber\\
&&-\kappa_c\cos\theta). \nonumber
\end{eqnarray}
The expressions for $R_{AB, AC,BC}$ are similar with the replacement $\kappa_o\rightarrow\kappa_o\bar n_o$. These coupling terms cause polariton correlation in the final steady state of the thermalization strokes and then result in work correlation in the isentropic strokes.  

In the case $\theta\sim\pi$ corresponding to the limit $gx\ll -G$, the correlations disappear due to the negligible values of correlation coefficients $R_{jk}$ and $S_{jk}$. Then the steady state of the three polariton modes are uncorrelated thermal states with the mean occupation numbers $\bar n_{A,B,C}$, respectively, and $\bar n_A\sim\bar n_b$ and $\bar n_B\sim\bar n_a$. For the opposite limit $gx\gg G$, the correlation is also near zero, $\bar n_A$ and $\bar n_B$ exchange their values. However, to other cases, the non-vanishing correlation will affect the steady population of the polariton modes. For example,  from the steady solution of the master equation (\ref{masterABC}) we have
\begin{eqnarray}
\langle\hat A^\dagger \hat A\rangle_s & = & \bar n_A-\frac{R_{AB}}{2\kappa_A}\langle\hat A^\dagger\hat B+\hat B^\dagger \hat A\rangle_s \nonumber\\
&&-\frac{R_{AC}}{2\kappa_A}\langle\hat A^\dagger\hat C+\hat C^\dagger \hat A\rangle_s.
\end{eqnarray}

The exact dependence of the populations and correlations on $\bar n_c$ and $G$ are obtained by the numerical simulation; specifically, for a large value of $G$ the RWA is invalid and the analytical form of the polariton transform is unavailable. The results are displayed in the Fig. 3 of the main text where we compared the exact work with that obtained by neglecting all coupling terms in the master equation (\ref{masterABC}) to show the influence of the correlation.


\begin{thebibliography}{99}

\bibitem{Abah2012} O. Abah, J. Ro{\ss}nagel, G. Jacob, S. Deffner, F. Schmidt-Kaler, K. Singer, and E. Lutz,  ``Single-ion heat engine at maximum power,'' Phys. Rev. Lett. {\bf 109}, 203006 (2012).

\bibitem{Rossnagel2016} J. Ro{\ss}nagel, S. T. Dawkins, K. N. Tolazzi, O. Abah, E. Lutz, F. Schmidt-Kaler, and K. Singer, ``A single-atom heat engine'', Science {\bf 352}, 325 (2016). 

\bibitem{Thomas2011} G. Thomas and R. S. Johal, ``Coupled quantum Otto cycle'', Phys. Rev. E {\bf 83}, 031135 (2011).

\bibitem{Altintas2015} See e.g. F. Altintas and \"O. E. M\"ustecapl\i o\v{g}lu, ``General formalism of local thermodynamics with an example:
Quantum Otto engine with a spin-1/2 coupled to an arbitrary spin'', Phys. Rev. E {\bf 92}, 022142 (2015), and references therein.

\bibitem{Brunner2012} N. Brunner, N. Linden, S. Popescu, and P. Skrzypczyk, ``Virtual qubits, virtual temperatures, and the foundations of thermodynamics
'', Phys. Rev. E {\bf 85}, 051117 (2012).

\bibitem{Fialko2012} O. Fialko and D. W. Hallwood,  ``Isolated quantum heat engine'', Phys. Rev. Lett. {\bf 108}, 085303 (2012).

\bibitem{Scully2003} M. O. Scully, M. S. Zubairy, G. S. Agarwal, H. Walther, ``Extracting Work from a Single Heat Bath via Vanishing Quantum Coherence'', Science {\bf 299}, 862 (2003).

\bibitem{Tercas2016} H. T{\c c}rcas, S. Ribeiro, M. Pezzutto, and Y. Omar, ``Quantum thermal mechanics fueled by vacuum forces'', Phys. Rev. E {\bf 95}, 022135 (2017).

\bibitem{Zhang2014a} K. Zhang, F. Bariani, and P. Meystre, ``Quantum optomechanical heat engine'', Phys. Rev. Lett. {\bf 112}, 150602 (2014).

\bibitem{Zhang2014b} K. Zhang, F. Bariani, and P. Meystre, ``Theory of an optomechanical quantum heat engine'', Phys. Rev. A {\bf 90}, 023819 (2014).

\bibitem{Mari2012} A. Mari and J. Eisert, ``Cooling by Heating: Very Hot Thermal Light Can Significantly Cool Quantum Systems'', Phys. Rev. Lett. {\bf 108}, 120602 (2012). 

\bibitem{Mari2015} A. Mari, A. Farace and V. Giovannetti, ``Quantum optomechanical piston engines powered by heat'', J. Phys. B: At. Mol. Opt. Phys. {\bf 48}, 175501 (2015). 

\bibitem{Elouard2015} C. Elouard, M. Richard and A. Auff\`eves, ``Reversible work extraction in a hybrid opto-mechanical system'', New J. Phys. {\bf 17}, 055018  (2015). 

\bibitem{Kurizki2015} D. Gelbwaser-Klimovsky and G. Kurizki, ``Work extraction from heat-powered quantized optomechanical setups'', Sci Rep. {\bf 5}, 7809 (2015).

\bibitem{Quan2007} H. T. Quan, Yu-xi Liu, C. P. Sun, and F. Nori, ``Quantum thermodynamic cycles and quantum heat engines'', Phys. Rev. E {\bf 76}, 031105 (2007).

\bibitem{Quan2009} H. T. Quan, ``Quantum Thermodynamic Cycles and Quantum Heat Engines (II)'', Phys. Rev. E {\bf 79}. 041129 (2009).

\bibitem{Huang2013} X. L. Huang, L. C. Wang, and X. X. Yi, ``Quantum Brayton cycle with coupled systems as working substance'', Phys. Rev. E {\bf 87}, 012144 (2013).

\bibitem{Dong2013} C. D. Dong, G. Lefkidis, and W. H\"ubner, ``Magnetic quantum diesel engine in Ni2'', Phys. Rev. B {\bf 88}, 214421 (2013).

\bibitem{Wu1998} F. Wu, L. Chen, F. Sun, C. Wu, and Yonghong Zhu, ``Performance and optimization criteria for forward and reverse quantum Stirling cycles'', Energy conversion and management {\bf39}, 733(1998).

\bibitem{Huang2014} X. Huang, X. Niu, X. Xiu, and X. X. Yi, ``Quantum Stirling heat engine and refrigerator with single and coupled spin systems'', Eur. Phys. J. D {\bf 68}, 32 (2014).

\bibitem{Andrews2014} See e.g. R. W. Andrews, R. W. Peterson, T. P. Purdy, K. Cicak, R. W. Simmonds, C. A. Regal, and K. W. Lehnert, ``Bidirectional and efficient conversion between microwave and optical light'' and the supplementary information therein, Nat. Phys {\bf 10}, 321 (2014).

\bibitem{noteG} The validity of the linearized optomechanical coupling description relies on the effective optomechanical couplings being much larger than the single-photon couplings, $|G_a|\gg|g_a|$ and $|G_b|\gg|g_b|$. In addition the stability of the linearized system requires all normal-mode frequencies to be positive, a condition that imposes an upper bound on $|G_{a,b}|$.

\bibitem{Wang2012a} Y. D. Wang and A. A. Clerk, ``Using Interference for High Fidelity Quantum State Transfer in Optomechanics'', Phys. Rev. Lett. {\bf 108}, 153603 (2012).

\bibitem{Wang2012b} Y. D. Wang and A. A. Clerk, ``Using dark modes for high-fidelity optomechanical quantum state transfer'', New J. Phys. {\bf 14}, 105010 (2012).

\bibitem{Tian2012} L. Tian, ``Adiabatic State Conversion and Pulse Transmission in Optomechanical Systems'', Phys. Rev. Lett. {\bf 108}, 153604 (2012).

\bibitem{Dong2012} C. Dong, V. Fiore, M. C. Kuzyk, and H Wang, ``Optomechanical Dark Mode'', Science {\bf 338}, 1609 (2012).

\bibitem{note1} To be more precise, the work extraction in isentropic strokes is independent on the phonon population, but the heat exchange in isochoriclike strokes includes the small contribution of the photon-phonon correlation.

\bibitem{Rogers2012} B. Rogers, M. Paternostro, G. M. Palma, and G. De Chiara, ``Entanglement control in hybrid optomechanical systems'', Phys. Rev. A {\bf 86}, 042323 (2012)

\bibitem{Bochmann2013} J. Bochmann, A. Vainsencher, D. D. Awschalom, and A. N. Cleland, ``Nanomechanical coupling between microwave and optical photons'', Nat. Phys {\bf 9}, 712 (2013).

\bibitem{Gerry2005} C. C. Gerry and P. L. Knight, ``Introductory Quantum Optics'', Cambrigde University Press, 2005.

\bibitem{footnoteW} This work operator approach is convenient to discuss the quantum variance and correlation with customary expressions. However its validity only holds up to second moment as discussed in Ref.~\cite{Engel2007}.

\bibitem{Engel2007} A. Engel and R. Nolte, ``Jarzynski equation for a simple quantum system: Comparing two definitions of work'', Eur. Phys. Lett. 79, 10003 (2007).





\bibitem{noteRWA} The rotating wave approximation and then the polariton transform Eq.~(4) are invalid for large $G$, so in the numerical calculation for Fig.~3 we retained the anti-rotating terms of the optomechanical coupling.

\bibitem{Garcia2016} M. Serra-Garcia, A. Foehr, M. Moler\'on, J. Lydon, C. Chong, and C. Daraio, ``Mechanical Autonomous Stochastic Heat Engine'', Phys. Rev. Lett. {\bf 117}, 010602 (2016).

\bibitem{Roulet2016} A. Roulet, S. Nimmrichter, J. M. Arrazola, and V. Scarani, ``Autonomous Rotor Heat Engine'', arXiv:1609.06011.


\end{thebibliography}
\end{document}